\documentclass[]{spieman}
\usepackage[utf8]{inputenc}
\usepackage{showyourwork}
\usepackage{xspace}
\usepackage{lineno} 
\usepackage{url}

\renewcommand{\baselinestretch}{1.0} 
\usepackage{amsmath,amsfonts,amssymb}
\usepackage{graphicx}
\usepackage{relsize}
\usepackage{acro}
\usepackage{listings}
\usepackage[colorlinks=true, allcolors=blue]{hyperref}
\definecolor{mygreen}{rgb}{0,0.6,0}
\definecolor{mygray}{rgb}{0.5,0.5,0.5}
\definecolor{mymauve}{rgb}{0.58,0,0.82}

\acsetup{
  single    = false,
  list/sort  = true,
  cite/group = true,
  cite/group/cmd = \cite,
  cite/group/pre = {\xspace},
}

\DeclareAcronym{prf}{
  short = PRF,
  long = Pixel Response Function
}

\DeclareAcronym{psf}{
  short = PSF,
  long = Point Spread Function
}

\DeclareAcronym{gpu}{
  short = GPU,
  long = Graphics Processing Unit
}

\DeclareAcronym{hmc}{
  short = HMC,
  long = Hamiltonian Monte Carlo,
  cite = {Betancourt2017}
}

\DeclareAcronym{mcmc}{
  short = MCMC,
  long = Markov Chain Monte Carlo,
  cite = {Metropolis1953}
}

\DeclareAcronym{gp}{
  short = GP,
  long = Gaussian Process,
  cite = {Rasmussen2005,Aigrain2022}
}

\DeclareAcronym{snr}{
  short = SNR,
  long = signal-to-noise ratio
}

\DeclareAcronym{pld}{
  short = PLD,
  long = Pixel Level Decorrelation,
  cite = {everest}
}

\DeclareAcronym{k2sff}{
  short = SFF,
  long = Self Flat-Fielding,
  cite = {self_flat-fielding}
}

\DeclareAcronym{cpm}{
  short = CPM,
  long = Causal Pixel Model,
  cite = {causal_pixel}
}

\DeclareAcronym{api}{
  short = API,
  long = Application Programming Interface
}

\DeclareAcronym{jit}{
  short = jit,
  long = Just-In-Time
}

\DeclareAcronym{opd}{
  short = OPD,
  long = Optical Path Difference
}

\newcommand\jax{\textsc{Jax}\xspace}
\newcommand\dlux{$\partial$Lux\xspace}
\newcommand\equinox{\texttt{equinox}\xspace}
\newcommand\zodiax{\texttt{zodiax}\xspace}
\newcommand\optax{\texttt{optax}\xspace}
\newcommand\numpyro{\texttt{numpyro}\xspace}

\title{Differentiable Optics with {\Large $\partial$}Lux: I - Deep Calibration of Flat Field and Phase Retrieval with Automatic Differentiation}

\author[a,*]{Louis Desdoigts}
\author[b,c]{Benjamin J. S. Pope}
\author[b]{Jordan Dennis}
\author[a]{Peter G. Tuthill}

\affil[a]{Sydney Institute for Astronomy, School of Physics, University of Sydney, NSW~2006, Australia}
\affil[b]{School of Mathematics and Physics, University of Queensland, St Lucia, QLD~4072, Australia}
\affil[c]{Centre for Astrophysics, University of Southern Queensland, West Street, Toowoomba, QLD~4350, Australia}

\begin{document} 
\maketitle

{\noindent \footnotesize\textbf{Send correspondence to L. Desdoigts: \\E-mail: \linkable{louis.desdoigts@sydney.edu.au }}}



\begin{abstract}
The sensitivity limits of space telescopes are imposed by uncalibrated errors in the point spread function, photon-noise, background light, and detector sensitivity. 
These are typically calibrated with specialized wavefront sensor hardware and with flat fields obtained on the ground or with calibration sources, but these leave vulnerabilities to residual time-varying or non-common path aberrations and variations in the detector conditions. It is therefore desirable to infer these from science data alone, facing the prohibitively high dimensional problems of phase retrieval and pixel-level calibration.
We introduce a new Python package for physical optics simulation, \dlux, which uses the machine learning framework \jax to achieve GPU acceleration and automatic differentiation ('autodiff'), and apply this to simulating astronomical imaging.
In this first of a series of papers, we show that gradient descent enabled by autodiff can be used to simultaneously perform phase retrieval and calibration of detector sensitivity, scaling efficiently to inferring millions of parameters.
This new framework enables high dimensional optimization and inference in data analysis and hardware design in astronomy and beyond, which we explore in subsequent papers in this series.

\end{abstract}

\section{Introduction}
\label{sec:intro}

At the vanguard of space-based astronomical imaging, photometry, and spectroscopy, imaging precision is limited by systematics introduced by aberrations in the optics and noise processes in the detector. While problems are ubiquitous in astronomy, we are motivated by several core examples. In exoplanet direct imaging, we want to achieve high resolution and high contrast simultaneously with instruments like JWST Coronagraphy \cite{Girard2022}, Aperture Masking \cite{Sivaramakrishnan2022}, or Kernel Phase \cite{Martinache2010,Kammerer2022} modes. The astrometric mission Toliman \cite{tuthill2018} aims to measure precise relative positions of the binary stars $\alpha$~Centauri~AB to reveal the gravitational influence of unseen planets; this will require micro-arcsecond, micro-pixel astrometric precision. We may also want to perform high precision photometry, whether with dedicated missions like \textit{Kepler} \cite{Borucki2010} and TESS \cite{Ricker2015}, or as ancillary science with Toliman. 

In each case, we face serious limitations from an imperfect knowledge of the \ac{prf} -- the map of the intra-pixel and inter-pixel variations in detector sensitivity; and also of the \ac{psf} -- the diffraction-limited pattern by which light from a point source like a star spreads across a detector.

In this series of papers, we present a new software package, \dlux, for fitting high-dimensional parametrized physical optics models to astronomical data. By using the Python library \textsc{Jax}\cite{jax}, we obtain \ac{gpu} hardware acceleration, as well as automatic differentiation or `autodiff' \cite{Margossian2018} features that enable high-dimensional optimization and inference with gradient descent or \ac{hmc}. In the present paper, we focus on the particular problem of estimating the \ac{prf} and \ac{psf} simultaneously - i.e. joint flat field calibration and phase retrieval. In Paper~II, we will show how autodiff enables improvements to hardware \textit{design}, directly calculating and optimizing the Fisher information and therefore fundamental figures of merit of an optical system end-to-end. And in subsequent papers, we will use this for end-to-end deconvolution of high angular resolution imaging with an unknown \ac{psf}. We make \dlux available as open-source software on GitHub\footnote{\href{https://github.com/LouisDesdoigts/dLux}{github.com/LouisDesdoigts/dLux}}, and encourage interested readers to use and contribute to this package as a community resource. This paper has been compiled using ShowYourWork \cite{Luger2021}, so a compiled version of this paper with all figures linking to the code used to produce them can be found here\footnote{\href{https://github.com/LouisDesdoigts/instrumental_calibration/tree/main-pdf}{github.com/LouisDesdoigts/instrumental-calibration/tree/main-pdf}}.

\subsection{Phase Retrieval}
The \ac{psf} depends on the successive planes through which light passes from the entrance pupil of the telescope through to the detector. This can be calculated from physical optics: for a simple camera bringing light from the pupil to focus at a detector plane (the regime of Fraunhofer diffraction), the \ac{psf} is the Fourier transform of the input wavefront. In the more general Fresnel regime where the detector is slightly out of focus, there are additional quadratic phase factors before and after this Fourier transform. In more complicated instruments like coronagraphs \cite{Bowler2016}, the light might pass through multiple re-imaged pupil and focal planes, being operated on in each. In either case, the trouble is that the \ac{psf} is distorted by unknown aberrations - distortions in the wavefront, which can be represented as a spatially-varying phase map across the optical plane. 

Phase retrieval is the problem of inferring these aberrations from data\cite{schechtman2014}, which is in general ill-posed \cite{barnett2020}: because of the Hermitian symmetry of the Fourier transform, and because we measure \textit{intensity} and not electric field in optical astronomy, there is a large space of aberrations that would generate the same intensity \ac{psf}. Fortunately, this space can be restricted to physically-realistic solutions and readily solved by algorithms such as the Gerchberg-Saxton algorithm \cite{gerchberg1972}, using ideas from compressed sensing \cite{candes2011}, or by machine learning \cite{metzler2018,isil2019,nishizaki2020}.  Phase retrieval was memorably performed to infer and correct the serious aberration on the \textit{Hubble Space Telescope} mirror at launch \cite{hubble_phase_ret}. 
Earlier work in the vein of this paper has shown that the phase retrieval problem can be efficiently solved even in the case of detector nonlinearity by taking a forwards model of physical optics, obtaining partial derivatives with autodiff, and optimizing its parameters by gradient descent \cite{jurling_fienup,phase_ret_and_design}, including a conference presentation of an early version of the present work \cite{Desdoigts2022}.

\subsection{Detector Calibration}

High-precision, high-cadence time series of space photometry from \textit{Kepler} \cite{Borucki2010}, K2 \cite{K2}, TESS \cite{TESS} and CHEOPS \cite{CHEOPS} space telescopes have been revolutionary for exoplanetary science \cite{Zhu2021} and stellar astrophysics \cite{Aerts21,Jackiewicz2021}. 

In many practical cases, photometric precision is limited by systematic errors due to variations in sensitivity within and between pixels. Changes in telescope pointing or variations in the \ac{psf} with focus couple to these inter- and intra-pixel variations in the flat field to produce changes in overall measured flux. This was particularly severe for the K2 mission where periodic thruster firings introduced a saw-tooth systematic on 6 hour timescales, comparable to the durations of exoplanet transits and asteroseismic signals of interest. 

Much work has been put towards developing data-driven self-calibration of the flat field in K2, which are of general interest and applicability for other space photometry missions. 
In \ac{k2sff}, the systematics are modelled with a spline regression of raw flux versus detector position, together with a linear combination of principal components of all light curves on the detector, and \textsc{k2sc} \cite{Aigrain2016} uses a \ac{gp} similarly. 
The \ac{cpm} constrains instrumental effects by measuring correlations in the light curves of spatially distant, causally-disconnected pixels.
In \ac{pld} the light curve is detrended by a linear combination of regressors formed by the ensemble of normalized pixel time series contributing to the light curve, together with their higher-order products.
Halo photometry \cite{halo, halo2} extracts light curves from a weighted sum of pixels where the weights are learned by minimizing the total variation of the resulting light curve, a convex optimization which is tractable because of autodiff.

Accurate measurement of background noise is also a serious issue for imaging science, where for example bad pixels, background noise, and uncertainties in the detector systematics limit sensitivity of aperture masking interferometry and kernel phase \cite{Kammerer2019}, and for low-surface-brightness science such as with \textit{Euclid}\cite{EuclidCollaboration2022}. In each case it is normal for detectors to be rigorously calibrated on the ground before flight, and with dedicated observations in space; but at the levels of contrast and resolution required for aperture masking on JWST (ideally $10^{-4}$ at a few $\lambda/D$) even small residual miscalibrations can seriously affect performance \cite{Sivaramakrishnan2022} and we require some method of self-calibration from science data themselves.

\section{\dlux Differentiable Optical Models}
\label{sec:dlux}

Rather than working with reduced data products like visibilities and light curves, an alternative is to directly model the pixel-level images with a parametrized model of the optics and detector. 
For example, the popular package \textsc{poppy} \cite{poppy} can accurately simulate Fresnel and Fraunhofer propagation, with a WebbPSF extension for preset JWST instrument models \cite{Perrin2014}. 
In principle, it is possible to fit such a model to data with \ac{mcmc}, fitting the positions of stars and a modal basis representation of aberrations; but common samplers can be very slow, or fail to converge in high dimensions \cite{Huijser2022}. This restraint is lifted by using \ac{hmc}, which requires computation of the gradient of the log-likelihood with respect to parameters. 

The core technology that enables this for physics also underpins the last decade's revolution in deep learning \cite{lecun15}: autodiff. By application of the chain rule, it is possible to calculate the partial derivatives of almost arbitrary numerical functions with respect to their floating-point arguments, so long as they are executed in an appropriate software framework. 
Considerable private-sector and open-source investment has gone into building autodiff frameworks to enable machine learning, such as Theano \cite{theano}, TensorFlow \cite{tensorflow2015}, PyTorch \cite{pytorch}, native support in the Julia language \cite{julia}, and the framework we apply in this paper, \textsc{Jax} \cite{jax}. \jax has an \ac{api} mirroring the common Python array library \textsc{NumPy} \cite{numpy}, while supporting \ac{gpu} acceleration, \ac{jit} compilation, and autodiff.

In this paper we introduce a new optical simulation package, \dlux, available under an open-source BSD 3 license. It significantly extends the capabilities of \texttt{morphine} \cite{pope2021,phase_ret_and_design}, a previous \jax optics package from our team based on \texttt{poppy}, being rewritten from the ground up taking full advantage of more-recent \jax features and libraries.  It is built in an object-oriented \jax framework called \zodiax which extends \equinox \cite{kidger2021equinox} for scientific programming. An optical system consisting of an \zodiax module wrapping a stack of `layers' which are themselves \zodiax modules applying phase or amplitude screens, or performing Fourier or Fresnel transforms on a wavefront. This allows us to re-frame the idea of a computational optical model by analogy to a `parametrized neural network', with each operation performed on the wavefront analogous to a neural network layer encoding the physics of the transformation. The object-oriented nature of \zodiax allows for larger and more complex models to be built and seamlessly integrated than would be possible within the purely functional framework default to \jax. \zodiax is also built to be directly integrated into the \jax framework, allowing for all \jax functions to seamlessly interface with any \zodiax models and vice-versa. We use the optimization library \optax \cite{optax2020github} for gradient-descent optimization, and the probabilistic programming library \numpyro \cite{Phan2019} for \ac{hmc}.

There are several other frameworks for automatically-differentiable optics simulation:
the most similar is WaveBlocks \cite{page2020}, a PyTorch package for object-oriented modelling of fluorescence microscopy. 
The DeepOptics package \cite{sitzmann2018} has been developed in TensorFlow for camera design and computational imaging, and
WaveDiff in TensorFlow has been used for detector calibration \cite{liaudat2021} and \ac{psf} modelling \cite{Liaudat2022}. 
We believe that \dlux fills a unique niche in this ecosystem as the most general modelling framework for physical optics. This has driven a design focus on flexibility and ease of open-source contribution and development. The goal is to provide a general object-oriented framework that enables a new user to harness the advantages of autodiff and hardware acceleration provided by \jax with its user-friendly \textsc{NumPy}-like API.


\section{Optical and Instrumental Modelling}
\label{sec:simulation}

\begin{figure*}
    \centering
    \includegraphics[width=\textwidth]{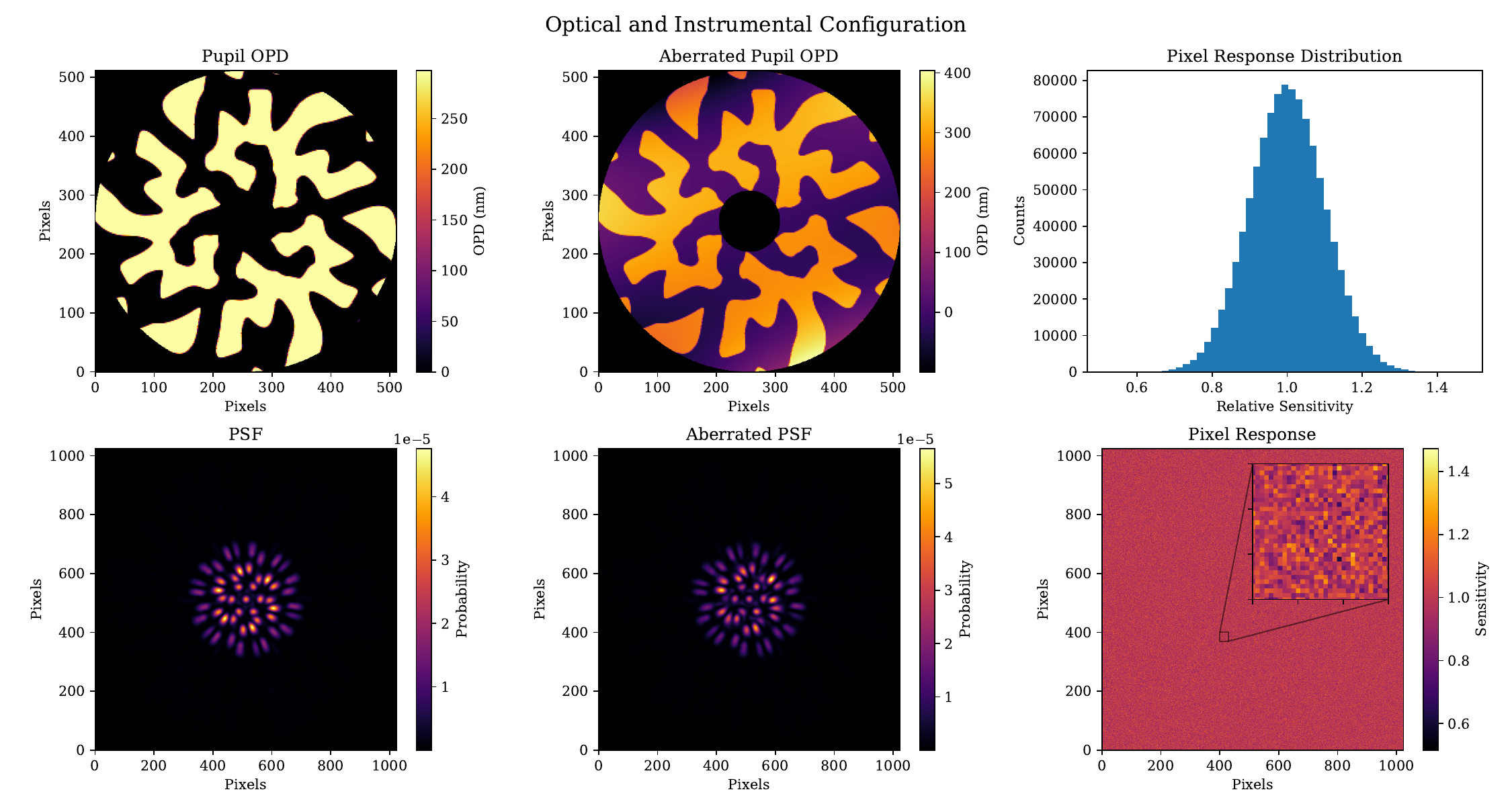}
    \caption{Summary of the optical configuration. Left panel: The top plot show the \ac{opd} of the pupil at the aperture of the telescope. The binary values create a half-wave step at the mean observation wavelength. The bottom plot shows the resulting large single-star \ac{psf} without aberrations applied. Middle panel: The top plot shows the total \ac{opd} of the pupil with the optical aberrations applied using low-order Zernike polynomials that is used to generate the data. The bottom plot shows the resulting single-star \ac{psf} with the aberrations applied. Clearly these aberrations have a large effect on the \ac{psf} and would make recovering information very difficult without appropriate calibration. Right panel: The top plot shows the histogram of the \ac{prf} that is applied in the focal plane. These values have a large spread and would greatly affect any results without calibration. The bottom plot shows the full \ac{prf} across the whole detector, with a small zoomed region used to show the fine detail that can not be seen when examining the full detector.}
    \label{fig:optics}
    \script{plot_optics.py}
\end{figure*}

Within the $\partial$Lux framework we can construct an optical model with only a pupil and focal plane like so:

\begin{lstlisting}[language=Python,frame=single]
import jax.numpy as np
import jax.random as jr
import dLux as dl
import dLux.utils as dlu

# Define wavelengths
wavels = 1e-9 * np.linspace(545, 645, 3)

# Basic Optical Parameters
diameter = 0.5 # meters
wf_npix = 256

# Construct a simple aperture
oversample = 3
coords = dlu.pixel_coords(oversample * wf_npix, diameter)
outer_circle = dlu.circle(coords, diameter / 2)
inner_circle = dlu.circle(coords, diameter / 10, invert=True)
transmission = dlu.combine([outer_circle, inner_circle], oversample)

# Construct Zernike aberrations
noll_inds = np.arange(4, 11)
coeffs = 2e-8 * jr.normal(jr.PRNGKey(1), (len(noll_inds),))
coords = dlu.pixel_coords(wf_npix, diameter)
basis = dlu.zernike_basis(noll_inds, coords, diameter)

# Detector Parameters
det_npix = 1024
det_pixsize = 12e-3  # arcseconds

# Build the layers
layers = [
    ("aperture", dl.layers.BasisOptic(basis, transmission, coeffs, normalise=True)),
    ("mask", dl.Optic(phase=np.load("mask.npy"))),
]

# Combine into Optics object
optics = dl.AngularOpticalSystem(wf_npix, diameter, layers, det_npix, det_pixsize)
\end{lstlisting}

The aperture object defines the pupil support as well as low-order phase errors parameterised by coefficients on Zernike polynomials. We include the first 4 radial terms, ignoring global piston, tip and tilt for a total of 7 terms. Coefficients were randomly drawn from a normal distribution of mean 0 and standard deviation 20\,nm.
These aberrations applied to the pupil mask and resulting \ac{psf} are shown in the middle panel of Figure~\ref{fig:optics}, where we see this induces a large amount of distortion in the \ac{psf}.

The mask object models the binary phase plate diffractive pupil from Toliman \cite{tuthill2018}, with a half-wave total \ac{opd} between the binary regions. We use this as an information-preserving way to spread the \ac{psf} across a large number of pixels which encodes the \ac{prf} information for a large number of pixels simultaneously. Alternative methods such as defocusing or using a diffuser \cite{Stefansson2017} come at the cost of eliminating Fourier information which constrain the wavefront and positions. An important consideration which is alleviated by the Toliman pupil is that focal-plane wavefront sensing suffers a sign ambiguity for all phase modes that are inversion-symmetric about the origin, and for unambiguous phase retrieval we therefore need an asymmetric or odd-mode symmetric pupil \cite{Martinache2013}. This pupil and corresponding aberrated \ac{psf} are shown in the left panel of Figure~\ref{fig:optics}.

\newpage

This is then propagated to the $1024\times1024$ pixel focal plane by a two-sided matrix Fourier transform \cite{Soummer2007,Martinache2020}. The detector \ac{prf} is modelled by multiplying the resulting \ac{psf} by a pixel sensitivity map which is drawn from a normal distribution with mean 0 and standard deviation 0.1. The histogram of pixel sensitivities and total pixel sensitivity map is shown in the right panel of Figure~\ref{fig:optics}. Here we show how to create this within the $\partial$Lux framework and create an instrument object to model this:

\begin{lstlisting}[language=Python,frame=single]
# Pixel response
pix_response = 1 + 0.1 * jr.normal(jr.PRNGKey(2), [det_npix, det_npix])

# Create Detector object
detector = dl.LayeredDetector([dl.ApplyPixelResponse(pix_response)])
\end{lstlisting}

This operation is then performed over 20 stars with uniform 100\,nm bandpass spectra sampled at three wavelengths 545\,nm, 595\,nm and 645\,nm. This total model requires the calculation of 60 individual \ac{psf}s over a $1024\times1024$ detector from a $512\times512$ pixel wavefront. This calculation is vastly dominated by the propagation of the wavefront using a 2-sided MFT, which requires two matrix multiplications for each individual \ac{psf} of sizes $[N_o \times N_i] \cdot [N_i \times N_i]$ and $[N_o \times N_i] \cdot [N_i \times N_o]$ where $N_i = 512$ is the input size of the wavefront and $N_o = 1024$ is the output size of the wavefront. This takes a total of $\sim$ 1.25 seconds to evaluate on an Apple M1 CPU, with all operations performed using FP32 precision default to \jax. FP32 was chosen as this problem does not require precision over a large dynamic range, however FP64 calculations can be performed as required. We also apply 4 integer-pixel dithers of +- 20 pixels in the (x,y) directions, equivalent to a 1 $\lambda/D$ dither for a total of 5 images with the total flux spread evenly across each image. This helps the over-parameterisation consequential from having a free parameter for each pixel. Using pixel-integer dithers similarly allows for a single PSF calculation on a slightly larger detector plane (1064x1064 in this case) and then sampling the smaller 1024x1024 regions to produce the full set of 5 images.

The data set is generated by taking this set of images and applying Poisson photon noise to the \ac{psf}s. Using this image we recover stellar positions and fluxes, optical aberrations, and individual pixel sensitivities. This gives us 40 stellar positional parameters (20 RA and DEC each), 20 stellar flux parameters, 7 optical aberration parameters, and 1,048,576 detector parameters: the dimensionality of the problem is vastly dominated by the \ac{prf} parameters, which is why this problem is made tractable only by using autodiff.


\section{Optimisation and Results: Phase Retrieval and Flat Field Estimation}
\label{sec:phaseretrieval}

\begin{figure*}
    \centering
    \includegraphics[width=\textwidth]{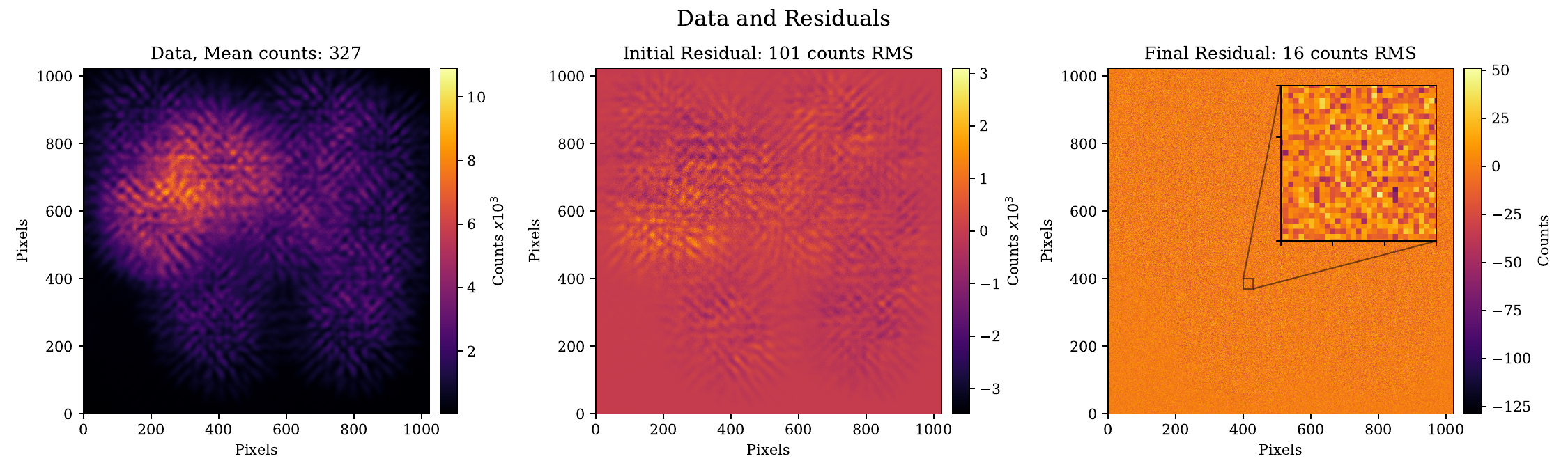}
    \caption{Left: The sum of the images from which the \ac{prf} is recovered. By eye it is clearly difficult to disentangle the astrophysical information. The large number of overlapping \ac{psf}s is chosen in order to spread light across the majority of the detector so that we encode the \ac{prf} information for as much of the detector as possible. Middle: The sum of the residual of the data and the initial uncalibrated model. Clearly these residuals are large, showing that there is a large amount of calibration required. Right: This same residual after the model has been optimised. The residual values are much smaller with no discernable structure remaining. A small zoomed region is shown so that the individual pixel-level residuals can be seen.}
    \label{fig:data_residual}
    \script{plot_data_resid.py}
\end{figure*}

To recover these parameters we initialise a naive unaberrated optical model with a uniform \ac{prf}. Stellar positions were perturbed by adding a random value drawn from a normal distribution mean 0 deviation 1 pixel and fluxes were perturbed by multiplying their value by a random value drawn from a normal distribution mean 1 deviation 0.1. We will use this model to recover the true values. Figure~\ref{fig:data_residual} shows the data and the initial and final residuals.



In order to recover these input aberrations, we perform gradient descent using Adam \cite{Kingma2014} as implemented in \optax \cite{optax2020github}, minimizing the posterior of a per-pixel Poisson log-likelihood with a $\chi^2$ log prior on the \ac{prf} values, matching the true distribution of mean 1 deviation 0.1. We take the gradient of this loss function with respect to the position and flux of all stars, the response in all pixels, and the Zernike coefficients. We also use a staged optimisation strategy, initially optimising the positions, fluxes and optical aberrations in order to recover a good \ac{psf} model before learning the \ac{prf}. This is necessary to circumvent the high degree of covariance that each individual pixel experiences with the parameters of the \ac{psf}. To achieve this the learning rate of the \ac{prf} is set to zero for the first 100 epochs of optimisation, after which the \ac{prf} learning rate is initialised while the position, flux and aberrations parameters are frozen for the remaining 50 epochs. This helps to circumvent the global covariance between the mean \ac{prf} and flux which can result in an overestimation of the flux parameters as the model tries to fit to the noise.

The optimisation is performed for a total of 150 epochs, with each value-and-gradient calculation taking $\sim$ 2.75 seconds to evaluate on an Apple~M1~Max CPU, for a total optimisation time of $\sim$ 6.5 minutes. Faster convergence can be achieved by careful tuning of the optimisation hyper-parameters such as learning rate and momentum, however we take a slower approach in this work in order to get good performance across a range of fluxes and PRFs which is analysed in Section~\ref{sec:noise}.

\begin{figure*}
    \centering
    \includegraphics[width=\textwidth]{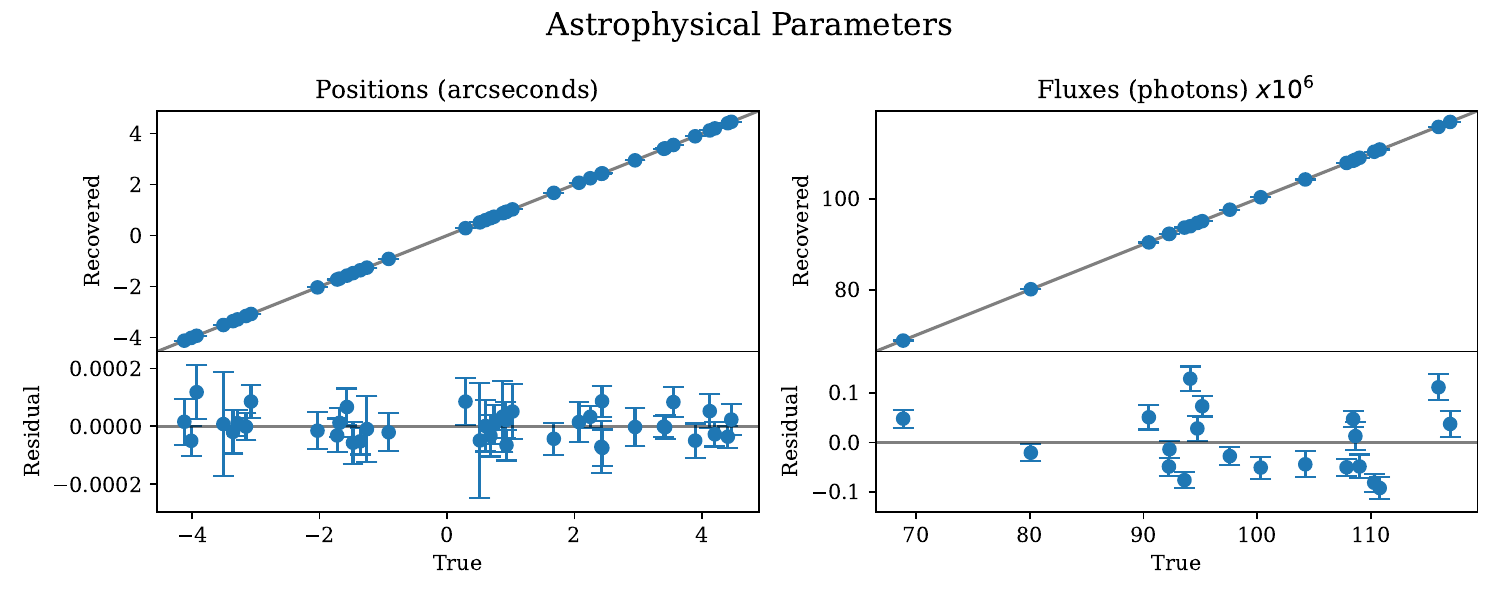}
    \caption{Left: recovery of the individual RA-Dec positions of each star, with the top section showing the correlation between the true and recovered values in units of arcseconds. The bottom section shows the individual residuals for each. Clearly these parameters have been well recovered through the optimisation. Right: the recovery of the flux parameters in units of photons. The top section shows the correlation between the true and recovered values, and the bottom showing the resulting residuals. The error bars are the 1-$\sigma$ deviations calculated from the covariance Hessian matrix marginalised over positions, fluxes and aberrations.}
    \label{fig:astro_params}
    \script{plot_astro_params.py}
\end{figure*}

\begin{figure*}
    \centering
    \includegraphics[width=\textwidth]{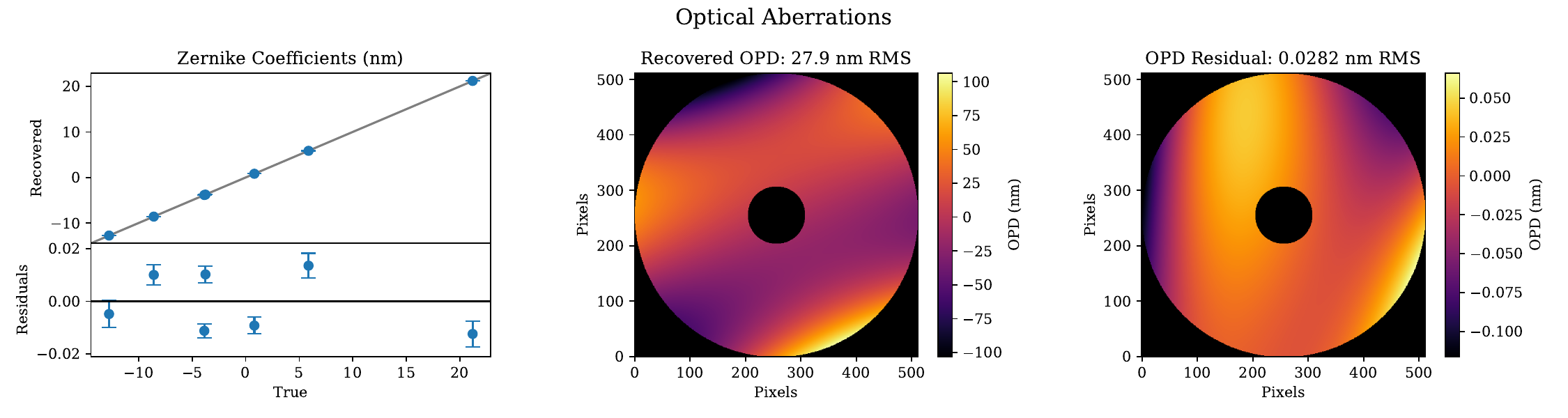}
    \caption{Recovery of the optical aberrations after optimisation. Left: the amplitudes of the Zernike polynomial coefficients, with the top section showing the correlation plot of the true and recovered values, and the bottom showing the individual residuals. Middle and right: the true total \ac{opd} and the residuals generated by these values respectively. The error bars are the 1-$\sigma$ deviations calculated from the covariance Hessian matrix marginalised over positions, fluxes and aberrations.}
    \label{fig:aberrations}
    \script{plot_aberrations.py}
\end{figure*}

This optimisation strategy performs well, with Figure~\ref{fig:data_residual} showing the final residual image with no visually discernable structure. A further analysis of parameter recovery over a range of fluxes and PRFs is shown in Section~\ref{sec:noise}. Correlations and residual plots of the positions, fluxes and aberrations after the optimisation are shown in Figure~\ref{fig:astro_params} and Figure~\ref{fig:aberrations}. Figure~\ref{fig:aberrations} shows the recovery of the optical aberrations, with a final residual aberration of 0.0282\,nm RMS.

\begin{figure*}
    \centering
    \includegraphics[width=\textwidth]{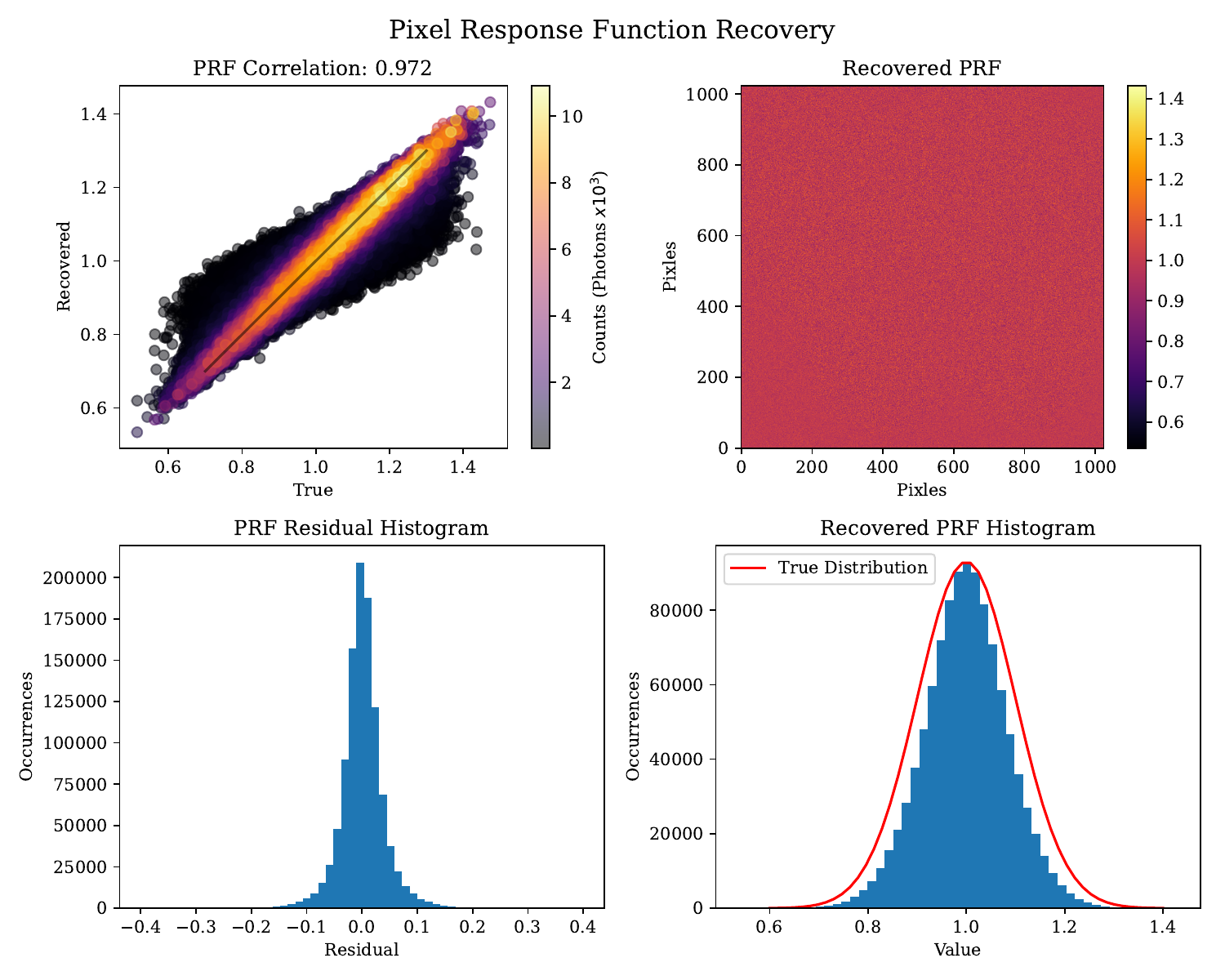}
    \caption{\ac{prf} values after optimisation. Left: Correlation between the pixel-level true and recovered values. Each point is color-coded with the total flux incident to each pixel, ie its \ac{snr}. The pixels with a greater signal are recovered better as expected. Right: Histogram of these residuals, with the majority of values being very well recovered and small symmetric about a residual of zero.}
    \label{fig:flat_field}
    \script{plot_FF.py}
\end{figure*}

Finally, Figure~\ref{fig:flat_field} shows a correlation plot of the true and recovered sensitivity values and the corresponding residual histogram. The points of the correlation plot are colour-coded with the total counts for each pixel, showing as expected that pixels with higher photon counts having a better recovery of the \ac{prf} since these pixels have a greater \ac{snr}. Overall the parameters are very well recovered down to the noise level.

We can also leverage the differentiability of these models in order to estimate the errors in these parameters under the Laplace approximation \cite{bard1974nonlinear}, by taking the inverse Hessian of the Poisson likelihood. Using this method marginalised errors can be calculated incorporating parameter covariance. There are some limitations using this method as it requires the calculation of the covariance matrix over all parameters - which in this model is over a million, giving a covariance matrix of over a computationally-intractable trillion values. Instead we calculate the errors marginalising \textit{only} over stellar positions, fluxes and optical aberrations, with pixel sensitivities held constant. This provides us with an accurate calculation of the true errors when photon noise is dominant, but it will remain an underestimate when the \ac{prf} is the dominant noise source. We chose to estimate our errors in this manner as opposed to the conventional photon-noise scaling relations in order to better capture covariance between parameters. In Section~\ref{sec:noise} we provide a comprehensive analysis of our results with respect to these photon-noise scaling relations.



The use of a forwards model clearly requires a good match between the true and model \ac{psf}. The benefit of using differentiable optical models however is that it provides the perfect platform upon which to directly learn this model from the data, with the ability to scale the complexity as required for each problem. Similarly since we are working with direct image data as opposed to reduced data-products, miss-calibrations of the system are simple to diagnose through direct examination of the residuals of the data and model. Figure~\ref{fig:data_residual} demonstrates that in this example we can be sure of a good fit because the final residual contains no visually identifiable structure.

\section{Noise \& Performance Analysis}
\label{sec:noise}

This section examines how well this method performs across a range of both \ac{prf} and flux values, which are our two dominant noise sources. Recovering parameters in a Poisson-noise-dominated regime for simple optical systems should have the following relationships:

$$\sigma_{\text{flux}} = 1/\sqrt{N_\text{phot}} \ \ \text{(photons)}$$

$$\sigma_{\text{position}} = \frac{1}{\pi} \sqrt{2 / N_{\text{phot}}} \frac{\lambda}{D} \ \ \text{(radians)}$$

$$\sigma_{\text{zernike}} = 1 / \sqrt N_{\text{phot}} \ \ \text{(RMS radians)}$$

Comparing the errors in the recovered parameters across a range of fluxes provides a good way to examine when we are in the photon noise dominated regime. After optimisation the errors are calculated for each parameter and compared to these expected values. ie: 

$$\sigma_{\text{relative}} = \sum_{i=1}^{N} \frac{1}{N} \frac{\sigma_{\text{i recovered}}}{\sigma_{\text{i expected}}}$$

\begin{figure*}
    \centering
    \includegraphics[width=1.0\textwidth]{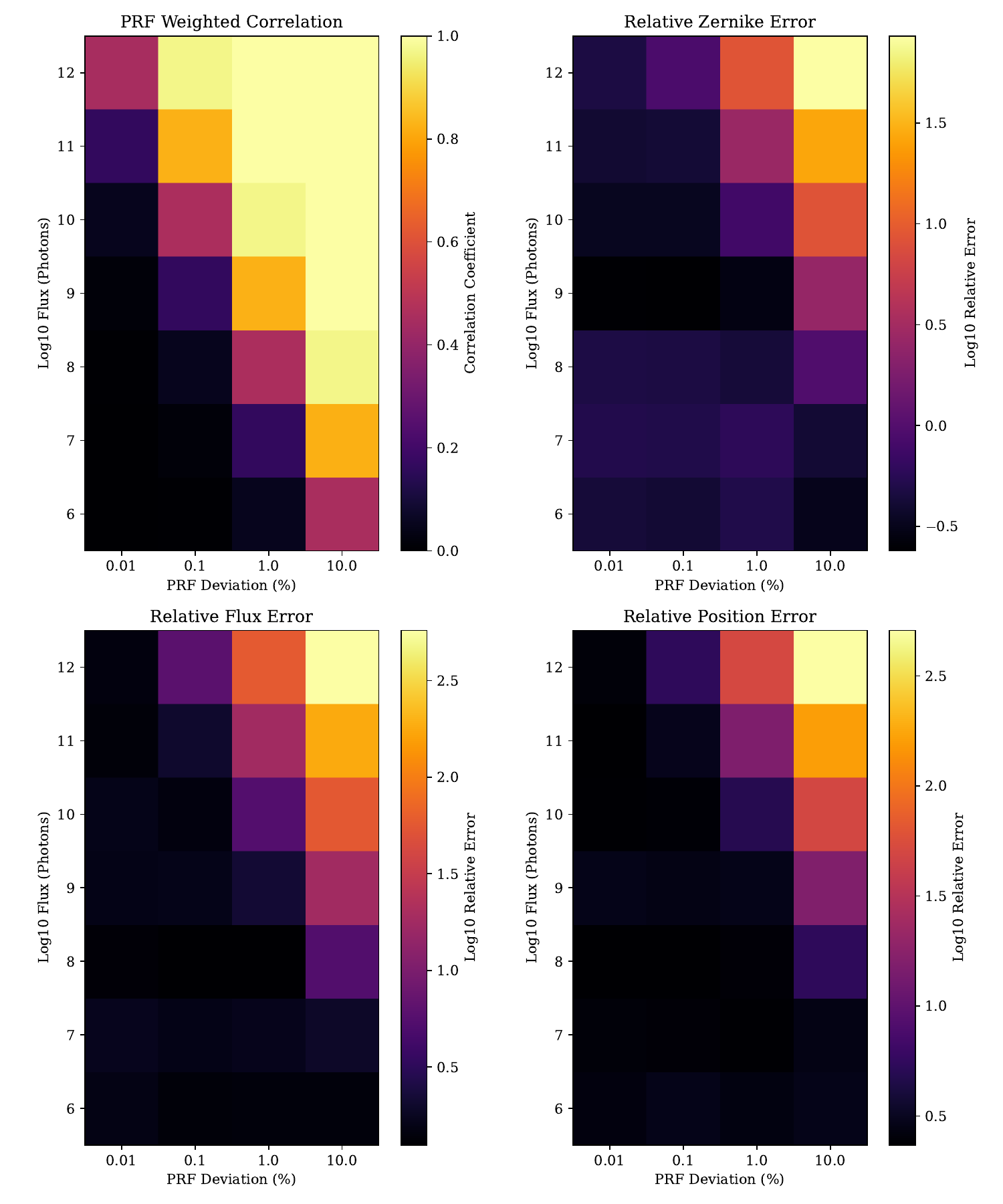}
    \caption{Recovered parameter error relative to photon noise expectation. Left: Mean relative per-mode Zernike error. Middle: Mean relative flux error. Right: Mean relative positional error.}
    \label{fig:noise}
\end{figure*}

The top-left panel of Figure~\ref{fig:noise} shows the weighted Pearson correlation coefficient~\cite{Costa2011} as a function of both flux and \ac{prf}. The transition where the \ac{prf} is recoverable denotes two regimes: one where photon noise dominates and another where \ac{prf} noise dominates. We have distinct results in these different regimes. Astrophysical and optical parameters are recovered to the photon noise expectation, while there is poor recovery of the \ac{prf} when photon noise is the dominant source and vice versa when the \ac{prf} is the dominant noise source. The top-right and bottom panels of Figure~\ref{fig:noise} show how $\sigma_{\text{relative}}$ changes as a function of both flux and \ac{prf}. These $\sigma_{\text{expected}}$ values are expected to be underestimates of the true values because they are not marginalised over all of the fitted parameters in the forwards model. By examining Figure~\ref{fig:noise}, we can see that there is a transition between two regimes: photon noise dominated and \ac{prf} noise dominated. In the photon noise dominated regime, we can see that the parameters are recovered to the level expected by these relationships and diverge from these once the \ac{prf} becomes the primary noise source.

It is important to note in this section that there is a degree of scatter in the recovered results. Due to the gradient descent optimisation strategy there is a degree of sensitivity to the optimisation hyper-parameters, such as learning rates which must be tuned based on the flux, momentum, and distance to the noise floor. These parameters are simple enough to tune for individual data sets with similar scales, but it is non-trivial to automate this process for data with a vast range of scales like those being analysed here. The hyper-parameters chosen were tuned on the 1e8 flux data set and roughly scaled to work with the other data sets, which is why we observe the best recovery for that set of parameters. These figures are designed to demonstrate the overall trends in the different noise regimes, rather than to be a precise analysis of the fundamental limits.

\section{Conclusions \& Future Work}
In this first paper of a planned series, we have demonstrated photometry, phase retrieval, and detector calibration using autodiff to make tractable gradient descent in very high dimensions. Our software \dlux integrates with a rich and user-friendly ecosystem of \jax packages for optimization and sampling, which point to a number of natural extensions.

The most obvious is to produce calibrated photometric time series, making use of correlations in time rather than the `bag of images' approach taken in this work. Instead of independently optimizing each frame, they could be treated hierarchically, with regularization imposed on each star's light curve. \ac{gp} models are widely used in astronomy, and \jax packages such as \texttt{tinygp} \cite{Aigrain2022} and GPJax \cite{Pinder2022} permit differentiable optimization and sampling that integrate efficiently with \dlux. 

Photometry of crowded fields is often of poor quality even from space telescopes, and there has been considerable investment in algorithms for deblending light curves in the K2 galactic plane campaigns \cite{Zhu2017} and in TESS with its large pixels \cite{Nardiello2019,Hedges2021,Higgins2022}. High dimensional optimization with autodiff in principle renders deblending tractable even close to the diffraction limit with unknown \ac{psf}s. Prior knowledge of a structured \ac{psf}, such as with the diffractive pupil architecture of Toliman \cite{Guyon2012,Guyon2013}, may further improve this deblending and retrieval. 

High-precision photometric measurements often need to also consider variations in the intra-pixel sensitivity. This work only examined the inter-pixel sensitivity for simplicity, however the \ac{psf}s examined here were sampled at 10x the Nyquist limit, well beyond the typical 1-1.5x Nyquist used in practice. The ability to recover sensitivity variations over these sub-fringe scales show that this method can be directly applied to recovery of both inter and intra-pixel sensitivity fluctuations.

High contrast imaging with the recently-launched JWST \cite{Gardner2006} and the upcoming Roman Space Telescope coronagraph \cite{Zellem2022} will in each case demand a new generation of data-driven \ac{psf} and detector calibrations beyond those used fruitfully on ground-based data \cite{Cantalloube2021}. While we have not attempted this in the present work, \dlux can flexibly model aberrations in the multiple intermediate planes of these coronagraphs, and in future work we intend to assess the practical and fundamental limits of multiplane phase retrieval and calibration to improve coronagraphic modelling.
Uncertainties in the flat field are also the leading contributor to noise in the exoplanet transmission spectroscopy with JWST \cite{Rustamkulov2022}, and a promising future direction with \dlux will be to build models not just of imaging data, but of spectrophotometry.

In Paper II in this series, we will tackle the isomorphic problem of hardware design. While we have tackled phase design in previous work \cite{phase_ret_and_design}, autodiff permits calculation and optimization of the Fisher information directly \cite{Coe2009}, which means we can optimize towards the fundamental limits of parameter sensitivity. In subsequent papers, we intend to explore parametric and nonparametric deconvolution of images at high angular resolution and contrast.

\section{Acknowledgements}

We thank Laurent Pueyo and Tim White for their helpful discussions.

We acknowledge and pay respect to the Gadigal people of the Eora Nation, upon whose unceded, sovereign, ancestral lands the University of Sydney is built; and the traditional owners of the land on which the University of Queensland is situated. We pay respects to their Ancestors and descendants, who continue cultural and spiritual connections to Country. 

This research made use of \textsc{NumPy} \cite{numpy}; Matplotlib \cite{matplotlib}; \textsc{Jax} \cite{jax}; \texttt{equinox} \cite{kidger2021equinox}; and \optax \cite{optax2020github}.

\bibliographystyle{spiebib} 
\bibliography{bib}

\end{document}